\documentstyle[preprint,revtex]{aps}
\begin{document}
\draft
\begin{title}
Role of Van Hove Singularities and Momentum Space Structure\\
in High-Temperature Superconductivity
\end{title}

\author{R. J. Radtke and K. Levin}
\begin{instit}
Department of Physics and the James Franck Institute,\\
The University of Chicago,
Chicago, Illinois  60637
\end{instit}
\author{H.-B. Sch\"uttler}
\begin{instit}
Department of Physics and Astronomy,
The University of Georgia,
Athens, Georgia  30602
\end{instit}
\author{M. R. Norman}
\begin{instit}
Materials Science Division,
Argonne National Laboratory,
Argonne, Illinois  60439
\end{instit}

\begin{abstract}

There is a great deal of interest in
attributing the high critical temperatures of the cuprates to either
the proximity of the Fermi level to a van Hove singularity or to structure of
the superconducting pairing potential
in momentum space far from the Fermi surface, the
latter being particularly important for spin-fluctuation-mediated
pairing mechansims.
We examine these ideas by calculating the critical temperature
$T_c$ for model
Einstein-phonon- and spin-fluctuation-mediated
superconductors within
both the standard, Fermi-surface-restricted Eliashberg theory and
the exact mean field theory, which accounts for the full momentum
structure of the pairing potential and the energy dependence of
the density of states.
Our calculations employ band structures chosen to model both the
$\rm La_2Sr_{2-x}CuO_4$ and $\rm YBa_2Cu_3O_{7 - \delta}$ families.
By using two models of spin-fluctuation-mediated pairing
in the cuprates, we
demonstrate that our results are independent
of the details of the dynamical susceptibility,
which is taken to be the pairing potential.
We also compare these two models against available
magnetic neutron scattering data, since these data provide
the most direct constraints on the susceptibility.
We conclude from our studies
that the van Hove singularity does not drastically
alter $T_c$ from its value when the density of states
is constant and that the effect of momentum structure is
significant but secondary in importance to that of the
energy dependence in the density of states.
\end{abstract}

\narrowtext

\section{INTRODUCTION}
\label{sec:intro}

The discovery of the cuprate superconductors with critical
temperates in excess of 90 K has motivated the
theoretical community to look for ways in which the accepted
BCS theory of superconductivity can yield such large
transition temperatures.
Accomplishing this task is particularly difficult because
the energy scale of both
phonons and spin fluctuations, which are prime candidates
for the pairing interaction in the
cuprates, are not so different from those
in the low-temperature superconductors.  Moreover, the
coupling strength of these interactions to the electrons implied by
transport measurements is small.
We focus on two hypotheses regarding this dilemma which
have attracted a great deal of attention in the literature.
The first asserts that structure in the
electronic density of states
near the Fermi level on the scale of the pairing boson energy is
mainly responsible for the enhanced critical temperature.
The second postulates that structure of the pairing potential
in momentum space far from the
Fermi surface is the dominant effect.

Investigations into the influence of
a strongly energy-dependent density of states on superconducting
properties has a long history
which begins with the A15 superconductors.
In these materials, the density of states is thought to exhibit
a square root singularity or
Lorentzian peak near the Fermi level, and
the analysis of the effect of this structure
on the critical temperature $T_c$ has proceeded
along both weak-coupling BCS \cite{FriedelA15,CarbotteWC} and
strong-coupling Eliashberg \cite{Horsch,CarbotteSC,Pickett} lines.
Recently, these approaches have been applied to the copper oxide
superconductors, where the density of states is thought to
contain logarithmic van Hove singularities due to the
quasi-two-dimensional nature of the $\rm CuO_2$ planes.
In the cuprates, however,
studies have been carried out mainly in the BCS limit
\cite{Hirsch,Lee,FriedelHTSC,Tsuei}.
These models result in
s-wave pairing and have been used to explain the isotope
effect in these materials.

Momentum structure in the
superconducting pairing potential has not been investigated
as systematically as has structure in the density of states.
One reason for the scarcity of work in this area may be that
the inclusion of momentum structure in the pairing potential
far from the Fermi surface calls into question the validity of
the usual Eliashberg theory and greatly complicates the calculation
of transport properties such as the resistivity.
For conventional, phonon-mediated superconductors, the difficulties
associated with wavevector-dependence of the pairing potential do
not arise, since Migdal's theorem \cite{Migdal} holds and effectively
restricts all quantities of interest to the Fermi surface.
For superconductivity produced by an electronic pairing mechanism,
in contrast, early studies have demonstrated that a simple RPA
treatment of the interaction in Eliashberg theory including
the full wavevector and frequency dependence of the pairing potential
overestimates $T_c$ \cite{Sham}.  In effect, the RPA form
of the electron-electron interaction is too strongly attractive
due to the neglect of vertex corrections.  In particular,
these vertex corrections are found to be important when the
characteristic pairing boson energy is larger than 5 \% of the
Fermi energy \cite{Sham}, which is a criterion easily satisfied by most
electronic pairing mechanisms.
More recent studies have indicated that the opposite conclusion may
hold in the two-dimensional Hubbard model:  vertex corrections are
significant and act to strengthen the electron-electron interaction
\cite{BS}.
In addition, since most transport calculations are based on an analogy
with the electron-phonon interaction and therefore use a version
of Migdal's theorem,
it is not clear whether or how these results can be used to describe
systems in which direct electronic interactions dominate.
The proper calculation of transport properties is important in the
theory of superconductivity, since the measured resistivity and
ac conductivity place constraints on the electron-pairing boson
coupling strength and hence on $T_c$ (see, for example,
Ref. \cite{RULN}).

Despite these difficulties, strongly wavevector-dependent
pairing potentials have attracted a great deal of interest,
especially within the context of spin-fluctuation-mediated
interactions.
Early work in this area was based on the Hubbard model treated
in the random phase approximation \cite{earlyAFSF}.
Current attention is centered around models of the cuprate
superconductors, in which antiferromagnetic spin fluctuations have been
shown to lead to d-wave paring states \cite{MBP,Ueda,RULN}.
For the cuprates, calculations of $T_c$ have been conducted in both
weak- and
strong-coupling Eliashberg formalisms and have used models
fit to the dynamical susceptibility obtained from
neutron scattering measurements \cite{RULN}, deduced from NMR
data \cite{MBP}, or based on
theoretical ideas regarding itinerant magnetism \cite{Ueda}.

In all cases, the wavevector dependence of the pairing potenital,
which is assumed to be given by the dynamical susceptibility,
is crucial to the existence of the superconducting phase; if
the wavevector dependence is removed from these models, $T_c$
vanishes.
The issues raised above regarding the applicability of
Eliashberg theory to situations with strong momentum structure in
the interaction are therefore extremely important here.
For the case of antiferromagnetic spin-fluctuation-mediated
interactions, work by Millis \cite{Millis} has indicated that
there may be a Migdal theorem which holds
for this type of pairing mechanism,
but this conclusion is controversial \cite{KS}.
For the present purposes, we put aside reservations
about Migdal's theorem and transport constraints on the coupling
constant in
order to examine the effect of momentum structure on $T_c$
within the Eliashberg formalism.  We do not directly answer the
question of whether the $T_c$ computed in this formalism has
anything to do with the physical transition temperatures.

To be specific, we will examine the ability of structure in the
density of states and the pairing potential to enhance
critical temperatures within Eliashberg theory.
In order to accomplish this goal, we compute $T_c$ for pairing
mediated by Einstein phonons and antiferromagnetic
spin fluctuations using two approaches.  The first employs
the standard, Fermi-surface-restricted approximations to the Eliashberg
theory, which assume a constant density of states and only include
variation of the pairing potential around the Fermi surface
\cite{AM,Scalapino}.
The second solves the full Eliashberg
equations at fixed band filling and therefore accounts for
the energy-dependent density of states
and the full wavevector dependence
of the pairing potential.
Since there is no wavevector dependence for the
Einstein-phonon-mediated interaction, comparing the solution of the
full Eliashberg equations to that of the Fermi-surface-restricted
equations for these phonons
illustrates the effect of an energy-dependent density
of states on $T_c$.
As noted above, the spin-fluctuation-mediated models have a strong
wavevector dependence in addition to a variation in the density of
states due to the band structure.
Since we are unable to separate these two effects directly, we use
the insights gained from the Einstein phonon calculation to extract
the relative contribution of these structures to $T_c$ through a
comparision of the solutions of the full and Fermi-surface-restricted
Eliashberg equations.  For the models we consider, it is the
density of states variation which provides the dominant influence
on $T_c$.

\section{THEORY}
\label{sec:theory}

\subsection{The Eliashberg Equations}
\label{sec:El}

We compute $T_c$ in the standard mean field
formalism \cite{AM,Scalapino} in which the electron
self-energy is solved self-consistently from the single-exchange graph.
In light of the discussion in the Introduction regarding the
validity of this approach for calculating $T_c$ in the cuprates,
we reiterate that the critical temperatures obtained from this
theory must not be taken too literally.
Given this caveat, the equations for the electron self-energy
in Matsubara space $\Sigma ({\bf k}, i \omega_n)$
can be written in the Nambu matrix notation \cite{AM} as
\begin{eqnarray}
\Sigma ({\bf k}, i \omega_n)
  & = & - \frac{T}{N} \sum_{ {\bf k'} m}
         g^2 P ({\bf k} - {\bf k'}, i\omega_n - i\omega_m)
         \, \tau_b G ({\bf k'}, i\omega_m) \tau_b
  \label{eq:El}
\end{eqnarray}
where $G ({\bf k}, i \omega_n)$ is the electron
Green's function which satisfies the Dyson equation
\begin{eqnarray}
G^{-1} ({\bf k}, i \omega_n)
  =  G_0^{-1} ({\bf k}, i \omega_n) - \Sigma ({\bf k}, i \omega_n)
  \label{eq:GF}
\end{eqnarray}
and $G_0 ({\bf k}, i \omega_n)$ is the bare propagator.
In these expressions, $N$ is the number of unit cells in the crystal,
$T$ is the temperature, and $g^2 P$ is the pairing potential.
$\tau_b$ is a Pauli
matrix which corresponds to $\tau_3$ if the pairing interaction
occurs through
the density-density channel (e.g., Einstein phonons)
and $\tau_0$ if the pairing occurs
through the spin-spin channel (e.g., spin fluctuations).
Throughout this paper, we set $\hbar = k_B = 1$.
In solving these
equations, we work at a fixed number of holes per site $n$, which is
determined by
\begin{equation}
n = 1 + \frac{1}{N} \sum_{{\bf k}n} \, e^{-i\omega_n 0+} \,
     {\rm Tr} \left[ \tau_3 G({\bf k}, i\omega_n) \right] . \label{eq:bf}
\end{equation}
We solve Eqs. (\ref{eq:El}) - (\ref{eq:bf}) as a function of
temperature to give $T_c$.

Recent work has shown that these equations can be solved directly
on lattices of small size (32 x 32, 64 x 64) \cite{MBP,Serene}.
Alternatively, when momentum structure in the pairing
potential can be ignored, as for Einstein phonons, these equations can
be solved without discretizing the Brillouin zone
\cite{CarbotteSC,Pickett}.  We have performed calculations for the
Einstein phonons in the latter scheme and those for the spin
fluctuations in the former; in the Einstein phonon case, the two
schemes give the same critical temperatures.

Most of the calculations in this paper are done in the
strong-coupling limit, where
the effects of the modification of the normal-state propagator
by the pairing interaction are included.
For completeness, we also solve for $T_c$ in the weak-coupling
limit, where the
normal-state propagator is not renormalized;
these solutions amount
to solving the BCS equation with a retarded interaction.

In addition to employing the exact Eliashberg equations
[Eqs. (\ref{eq:El}) - (\ref{eq:bf})] to compute $T_c$, we also use
an approximate form of these equations
in which all the wavevectors and energies are restricted to the Fermi
surface.
For the Einstein phonon case, these equations are the
usual form of the strong-coupling Eliashberg equations \cite{AM};
the equations used
for the spin fluctuations have been described in an earlier
publication \cite{RULN}.
Throughout this paper, we will refer to the solution
of Eqs. (\ref{eq:El}) - (\ref{eq:bf}) as the exact Eliashberg
solution, and we will refer to the solution of the approximate
equations as the Fermi-surface-restricted solution.

\subsection{Model Interactions}
\label{sec:bosons}

In order to solve Eqs. (\ref{eq:El}) - (\ref{eq:bf}) in either
the exact or Fermi-surface-restricted scheme, two ingredients
are required:  the pairing potential $g^2 P$ and the band structure
$\epsilon_{\bf k}$.  In this section we will discuss the
model pairing potentials used in our calculations, paying particular
attention to the spin-fluctuation-mediated interactions and
comparing their spectral functions to the trends extracted
from available neutron data.
In the next section, we will describe the band structure.

In spin-fluctuation-mediated superconductors, the pairing potential
is taken to be the magnetic susceptibility.
We consider two models of this
susceptibility taken from Ref. \cite{RULN}
and Ref. \cite{MBP}, which will be referred to as the RULN and
MMP models, respectively \cite{SF1,MMPnote}.
The RULN model is a phenomenological fit to magnetic neutron
scattering measurements on $\rm YBa_2Cu_3O_{6.7}$ \cite{neutrons}
and simultaneously provides
a reasonable fit to the microscopic calculations of Ref. \cite{Qimiao}.
The de-oxygenated material is used in this model because
there is, as yet, no consensus between different
neutron scattering experiments regarding the behavior of the
fully oxygenated system.
Since the de-oxygenated material is nearer to the
antiferromagnetic insulator than the fully oxygenated compound,
we view the $\rm YBa_2Cu_3O_{6.7}$ susceptibility, and hence the RULN model,
as an overestimate of the strength of
the spin fluctuations in $\rm YBa_2Cu_3O_7$.
As opposed to the RULN model, the MMP model for the dynamical
susceptibility is inferred from
NMR data in the $\rm YBa_2Cu_3O_{7 - \delta}$ family; details
of this model may be found in Ref. \cite{MPT}.

The dependence of the spectral functions of these
two model interactions on frequency, temperature, and wavevector are shown
in Figs. \ref{fig:omega} - \ref{fig:q}.
{}From Fig. \ref{fig:omega}(a), we see that the spectral function of the
RULN pairing potential at the antiferromagnetic wavevector
$(\pi,\pi)$
has a peak in frequency around 30 meV at low temperatures which
broadens and shifts to slightly larger frequencies at higher
temperatures.
For comparison, neutron scattering
data \cite{inset1} are shown in the inset to Fig. \ref{fig:omega}(b) and
demonstrate that the experimentally measured spin
fluctuation frequency is around 30 to 40 meV.
This spin fluctuation frequency is found to be roughly
constant or slightly decreasing with temperature.
The corresponding plot for the MMP model is displayed in
Fig. \ref{fig:omega}(b) and
shows a similar temperature dependence
of the peak frequency and peak width,
but the magnitude of the peak frequency, roughly
5 to 10 meV, is smaller than that observed experimentally
at $(\pi,\pi)$ .
We note, however, that the spin fluctuation energy in the MMP model
is strongly wavevector-dependent and can be as large at 1000 meV
at the zone center at low temperatures \cite{saturation}.

The temperature dependence of these model interactions is compared in
Fig. \ref{fig:t}, and experimental data are presented in the inset
to Fig. \ref{fig:t}(b) \cite{inset2}.
We see that the spectral function of the
RULN pairing potential shows a strong temperature dependence
at high frequencies
which disappears at lower frequencies, in agreement with the
data.
The MMP model has the opposite tendency.  The low-temperature behavior
of the two models is also different:  the MMP spectral function
always exhibits a
downward curvature for the parameters shown, but the
low-temperature curvature of the RULN interaction changes sign at high
frequencies.

Finally, we compare the wavevector-dependence of these models
in Fig. \ref{fig:q} and include experimental data in the inset to
Fig. \ref{fig:q}(b) \cite{inset2}.
In both cases, the spectral function has maxima at the $(\pi,\pi)$
points of the first Brillouin zone.  The RULN model displays a
constant half-width of this maximum and an increasing peak height
as a function of increasing frequency for the frequencies shown.
The MMP model shows a
weakly increasing half-width but a decreasing peak height as a
function of increasing frequency.
The measured full width at half-maximum $\Delta q$
is approximately 0.2 r.l.u.
for $\rm YBa_2Cu_3O_{6.7}$ and increases to roughly 0.3 r.l.u. in
$\rm YBa_2Cu_3O_{7}$ \cite{inset1}.
In the MMP model at the latter stoichiometry,
$\Delta q$ is about a third of the experimental value;
in the RULN model for
$\rm YBa_2Cu_3O_{6.7}$, $\Delta q$ is roughly 0.2 r.l.u.

Since the RULN model was constructed to describe the neutron scattering
experiments, it is not at all surprising that it accounts for those
measurements better than the MMP model, which evolved from fits to NMR data.
If we now consider the results from NMR experiments, we see that
the MMP model provides good fits to these results \cite{MMP,MPT}.
In contrast, the RULN model agrees with the Cu NMR
relaxation rate but does not yield the form factor cancellations
which are necessary to explain the observed Korringa behavior
at the O sites in the fully oxygenated system.
It may be inferred from the inset to Fig. \ref{fig:q}(b)
that there is considerable weight in the magnetic susceptibility
away from $(\pi,\pi )$ in the de-oxygenated case, and
this situation
only intensifies as one moves towards $\rm YBa_2Cu_3O_7$.
If one makes the standard
assumption that the O nuclei relax via the same susceptibility
as do the Cu,
then it is difficult to reconcile this experimental fact with the
perfect form factor cancellations used in interpreting NMR data.
Thus, the inability of the RULN model to explain O site NMR
measurements implies more about an
inherent inconsistency between the neutron scattering and NMR results
than about either model.

If these experiments are in conflict, then
we must decide which experiment provides more reliable constraints on
the pairing potential.
In order to extract the wavevector dependence of the
magnetic susceptibility from NMR data, one must examine the form
factors which are used to compute the relaxation rate.
Neutron scattering, on the other hand,
measures this structure directly.  Additionally,
the frequencies important for forming Cooper pairs are those
probed by neutron scattering and not the low frequencies observed
in NMR.
Hence, it is reasonable to expect that
any pairing mechanism involving spin fluctuations which purports to
describe superconductivity in the cuprates must reproduce at least
the qualitative features of the magnetic neutron scattering
experiments.

Before proceeding, we note two commonalties in the two
theoretical models.
First, the order
parameter arising from spin-fluctuation-mediated interactions
with a spectral function peaked at the antiferromagnetic points
of the Brillouin zone must
have d-wave symmetry \cite{MBP,Ueda,RULN}.
Second, we observe that the average frequency of the spin fluctuations
is well
within the range of phonon energies in the high-$T_c$ cuprates
(roughly 0 - 80 meV) near the $(\pi,\pi)$ point, where these
pairing potentials are largest.

In addition to spin-fluctuation-mediated interactions, we also
employ an Einstein-phonon-mediated interaction.
The pairing potential in this case is simple;
it has no wavevector dependence and the
spectral function has a delta function peak at the
phonon frequency $\omega_0$.  We choose $\omega_0$ to be 35 meV in order
to give the same energy scale as in neutron scattering
experiments.
Finally, we note that the resulting superconducting order parameter
in this case is required to have s-wave symmetry.

When applying the model interactions discussed in this section,
we should in principal use
the measured transport properties to constrain the
electron-pairing boson coupling constant as was done in Ref.
\cite{RULN}.  We abandon this constraint here and use
large coupling constants
in order to obtain critical temperatures which are
accessible to our numerical routines.  Consequently, the computed
critical temperatures are larger than what we view as realistic.

\subsection{Band Structure}
\label{sec:BS}

The electron energy dispersion we use is a two-dimensional,
tight-binding band structure which is chosen to agree with
local density approximation calculations and
angle-resolved photoemission experiments where they are available.
We work in the hole picture and write the dispersion as
\begin{equation}
 \epsilon_{\bf k} = - 2t_1 \left[ cos \,k_x a + cos \,k_y a
     + 2 t \,cos \,k_x a \,cos \,k_y a \right] - \mu.  \label{eq:BS}
\end{equation}
In this expression, we choose $t_1$ = 80 meV in order
to agree with the measured plasma frequency in
${\rm YBa_2Cu_3O_{6.7}}$ \cite{RULN,Orenstein}.
The parameter $t$ controls the Fermi surface rotation and is set
to 0.45 to model the ${\rm YBa_2Cu_3O_{7 - \delta}}$ (YBCO)
family and to 0.00 in order to model the
${\rm La_{2 - x}Sr_xCuO_4}$ (LSCO) family.
In both cases, the chemical potential
$\mu$ is determined from the constraint equation Eq. (\ref{eq:bf}).
We note that this band structure has a van Hove singularity in the
bare density of states when $\mu = 4 t_1 t$.

\section{RESULTS AND DISCUSSION}
\label{sec:results}

\subsection{Einstein Phonons}

Having discussed the theoretical framework underlying our
calculations, we now present our numerical results.
In Fig. \ref{fig:Ein}, we plot the $T_c$ of an Einstein-phonon-mediated
superconductor as a function of the band filling.
In this figure, we include the solutions to the strong-coupling
Fermi-surface-restricted, strong-coupling exact, and
weak-coupling exact equations for both the YBCO and LSCO band
structures \cite{Ein}.
One sees immediately that the
sharp peak in the Fermi-surface-restricted calculations caused by
the van Hove singularity is completely removed when account
is taken of the energy-dependent density of states
by the exact solution of the Eliashberg equations.
We emphasize that the Einstein phonon pairing potential
has no wavevector dependence,
so the difference between the Fermi-surface-restricted and exact
critical temperatures is due entirely to the energy-dependent
density of states.

Similar behavior has recently been observed by
Penn and Cohen \cite{Penn}, who used an effective density of states
smeared out by the inclusion of lifetime effects in a weak-coupling
computation.  Our
results may be understood within the same context, but with
$T_c$ playing the role of the lifetime; that is, the structure
in the density of states is removed by thermal broadening.
We also note that the weak-coupling exact calculation gives
a stronger $T_c$ enhancement near the van Hove singularity
than that produced from the
strong-coupling calculation.  Thus, the inelastic scattering due to
the electron-pairing boson interaction acts with temperature to
reduce the effect of sharp structure in the density of states on $T_c$.

Fig. \ref{fig:Ein} also illustrates that, for band fillings far from the
van Hove singularity, the Fermi-surface-restricted and
strong-coupling exact
calculations give nearly the same $T_c$, indicating that the standard,
Fermi-surface-restricted theory gives accurate results when there is
no strong variation in the electronic density of states.  Additionally,
we observe that the critical temperature in the exact calculations
goes to zero at the band edge (cf. Fig. \ref{fig:Ein}(a)),
as one expects when the number
of charge carriers vanishes.  The Fermi-surface-restricted $T_c$,
however, produces an unphysical non-zero value at the band edge.

In attempting to evaluate the effect of the van Hove singularity
on $T_c$, what one wishes to know is how the singularity affects
$T_c$ compared to that of an energy-independent density of states.
One can crudely estimate this flat density-of-states $T_c$ by
looking for the average value of the Fermi-surface-restricted
critical temperature in Fig. \ref{fig:Ein}.  This number can then be compared
to the $T_c$ computed in the exact calculation at the band filling
where the van Hove singularity is located.  In this way, we
estimate from Fig. \ref{fig:Ein}(a) that an energy-independent density of
states would give $T_c \approx$ 90 K, whereas the exact calculation
gives $T_c \approx$ 80 K at the van Hove singularity.  Similar
numbers can be extracted from Fig. \ref{fig:Ein}(b).  On the basis of this
analysis, we conclude that the van Hove singularity does not
strongly enhance $T_c$ and may even suppress it, in disagreement
with the work of Ref. \cite{Tsuei}.
This result supports a statement which
appeared several years ago:  $T_c$ alone does not provide a unique
indicator of structure in the density of states, since it is
determined only by a particular average of the
density of states \cite{Pickett}.
We also note that recent calculations of the interacting density
of states above $T_c$ show that the van Hove singularity in the
bare density of states is completely removed by interactions which
are strong enough to yield critical temperatures greater than
about 30 K \cite{Zhong}.  These calculations are completely
consistent with the featureless $T_c$ vs. band filling curves
which we report here and strengthen our conclusion that van
Hove singularities alone cannot explain the large critical
temperatures of the cuprates.

\subsection{Antiferromagnetic Spin Fluctuations}

Turning now to spin-fluctuation-mediated pairing, we show in
Fig. \ref{fig:RULN} the $T_c$ calculations using the RULN model
\cite{RULN}
for both the YBCO and LSCO band structures \cite{SF2}.
As in the s-wave case, the sharp peak in $T_c$
due to the van Hove singularity in the Fermi-surface-restricted
formalism is removed in the exact calculation.
In contrast to the s-wave case, though, the
Fermi-surface-restricted and exact $T_c$'s differ away from
the van Hove singularity
with the magnitude of the discrepancy
depending on the band filling.
In particular, we see that the exact Eliasbherg
$T_c$ is not necessarily larger than the Fermi-surface-restricted $T_c$.
In interpreting this behavior, one must keep in mind that the
spin fluctuation model has both wavevector dependence and energy
variation in the density of states which influence the results.
Comparing Fig. \ref{fig:RULN} to Fig. \ref{fig:Ein},
where only density of states
structure affects $T_c$, we see that the disappearence of the
sharp van Hove feature in the exact calculation in
Fig. \ref{fig:RULN} can be
attributed to the inclusion of density of states variation
in that scheme.
On the other hand, the discrepency between the exact and
Fermi-surface-restricted Eliashberg $T_c$'s away from the
van Hove singularity is most probably due to momentum
structure in the pairing potential.  Of the two effects, it is
clear the the density of states plays the dominant role in
causing the disagreement between the Fermi-surface-restricted
and exact Eliashberg critical temperatures.

In Fig. \ref{fig:MMP}, we plot $T_c$ vs. band filling
for the MMP model \cite{MBP} computed for the YBCO band structure
\cite{MMP1}.
We note that the qualitative features of the these curves
are the same as those noted above for the RULN model.
Hence, our results for spin-fluctuation-mediated superconductors
are not strongly model-dependent \cite{fixedZ,incom}.

\section{CONCLUSIONS}

To sum up, we have recovered the expected results that
Fermi-surface-restricted calculations should be
reliable when there is no sharp structure in the density of
states, when momentum structure in the pairing potential is not
critical, and when one is not near the band edge.
More importantly, we can conclude that the proximity of the
Fermi level to a van Hove singularity cannot substantially enhance
$T_c$ in either s- or d-wave superconductors.
When the energy dependence in
the density of states and strong-coupling effects are
properly taken into account, we find that the van Hove singularity
produces a broad peak in $T_c$ as a function of band filling but
does not increase $T_c$ markedly over the value computed from a
structureless density of states.
We also find that including the full wavevector dependence of the
pairing potential over the entire Brillouin zone is
important for computing d-wave transition temperatures in the
Eliashberg formalism but that the influence of this structure
off the Fermi surface on $T_c$
is less important than density-of-states effects.
In other words, there is a discrepancy between the exact evaluation
of the Eliashberg equations and the evaluation of the standard,
Fermi-surface-restricted equations, but the discrepancy is not
always large or positive and can be attributed mainly to the
effects of a strongly energy-dependent density of states.
Finally, we observe that these conclusions appear to be
model-independent.

Our results for spin-fluctuations must be treated cautiously.
As discussed in the Introduction, critical temperatures computed
in the Eliashberg theory for strongly momentum-dependent interactions
are not necessarily physical due to the neglect of vertex corrrections.
It is known in $^3$He that, although the ratio of the paramagnon
energy to the Fermi energy is small, Migdal's theorem still fails
\cite{Hertz}.  It is not clear whether a similar result holds for
antiferromagnetic spin fluctuations (anti-paramagnons) in the cuprates;
results have appeared in the literature which argue both points
of view \cite{Millis,KS}.
Additionally,
BCS theory was built up from the supposition that
only interactions near the Fermi surface are important for
driving the superconducting instability.  Whether or not this
theoretical underpinning must be re-examined if the pairing potential
is attractive over the entire zone has not been addressed.
Therefore, we view the results of this
paper pertaining to spin fluctuations
and all related work in the literature thus far as preliminary.

\acknowledgements

This work was supported by NSF-STC-9120000.
HBS would also like to acknowledge financial support from
NSF-DMR-8913878 and NSF-DMR-9215123 and computer support from
the University of Georgia; MRN was supported
by the U.~S.~Department of Energy,
BES-Materials Sciences, under Contract \#W-31-109-ENG-38.

\figure{Spectral function Im $\chi ({\bf Q}, \omega)$
in arbitrary units as a function
of frequency $\omega$ in meV for (a) the RULN (Ref. \cite{RULN}) and
(b) the MMP (Ref. \cite{MBP}) models of spin-fluctuation-mediated
interactions $\chi ({\bf q}, \omega)$ in $\rm YBa_2Cu_3O_{7-\delta}$.
The spectral functions are
evaluated at the antiferromagnetic
wavevector ${\bf Q} = (\pi,\pi)$ and temperatures T = 0 K (solid
line), 100 K (long-dashed line), 200 K (short-dashed line), and
300 K (dot-dashed line).
Inset to the second figure:  experimental data from Ref. \cite{inset1}
of $\rm YBa_2Cu_3O_{6.6}$ at T = 100 K plotted as in the main figures.
\label{fig:omega}}

\figure{Spectral function Im $\chi ({\bf Q}, \omega)$
in arbitrary units as a function
of temperature T in K for (a) the RULN (Ref. \cite{RULN}) and
(b) the MMP (Ref. \cite{MBP}) models of spin-fluctuation-mediated
interactions $\chi ({\bf q}, \omega)$ in $\rm YBa_2Cu_3O_{7-\delta}$.
The spectral functions are
evaluated at the antiferromagnetic
wavevector ${\bf Q} = (\pi,\pi)$ and frequencies $\omega$ = 10 meV
(solid
line), 20 meV (long-dashed line), and 30 meV (short-dashed line).
Inset to the second figure:  experimental data from Ref. \cite{inset2}
of $\rm YBa_2Cu_3O_{6.6}$ plotted as in the main figures for $\omega$ = 5 meV
(open circles),
8 meV (crossed boxes), 12 meV (solid circles),
and 16 meV (open diamonds).
\label{fig:t}}

\figure{Structure function
$S ({\bf q}, \omega) =
2 \,{\rm Im} \chi ({\bf q}, \omega)\, /\,(1 - e^{-\omega/T})$
in arbitrary units as a function
of wavevector ${\bf q} = (k,k)$ in reciprocal lattice
units for (a) the RULN (Ref. \cite{RULN}) and
(b) the MMP (Ref. \cite{MBP}) models of spin-fluctuation-mediated
interactions $\chi ({\bf q}, \omega)$ in $\rm YBa_2Cu_3O_{7-\delta}$.
The spectral functions are
evaluated at frequencies $\omega$ = 10 meV (solid
line), 20 meV (long-dashed line), and 30 meV (short-dashed line).
Inset to the second figure:  experimental data from Ref. \cite{inset2}
of $\rm YBa_2Cu_3O_{6.6}$ at T = 10 K and $\omega$ = 8 meV plotted
as in the main figures.
\label{fig:q}}

\figure{Critical temperature $T_c$ in K vs. the number of holes
per site $n$ in the
two-dimensional, tight-binding band structure of Eq. (\ref{eq:BS})
with $t_1$ = 80 meV and with
(a) $t =$ 0.45 to model the YBCO family
and (b) $t$ = 0.00 to model the LSCO family.  The pairing boson is
an Einstein phonon of frequency 35 meV.
$T_c$ is computed in the strong-coupling exact
(solid line), the weak-coupling exact
(dot-dashed line), and the strong-coupling Fermi-surface-restricted
Eliashberg schemes
(dashed line).  The sharp feaure in the Fermi-surface-restricted
calculation is the van Hove singularity.
Note that the electron-phonon coupling constant in the weak-coupling
calculation is half that used in the strong-coupling calculations
and that
the coupling constants, and hence the $T_c$'s, are not necessarily
physical (see text).\label{fig:Ein}}

\figure{Critical temperature $T_c$ in K vs. the number of holes
per site $n$ in the
two-dimensional, tight-binding band structure of Eq. (\ref{eq:BS})
with $t_1$ = 80 meV
and with (a) $t =$ 0.45 to model the YBCO family
and (b) $t$ = 0.00 to model the LSCO family.
The pairing is mediated by spin-fluctuations as described in
Ref. \cite{RULN}.
$T_c$ is computed both in the strong-coupling
exact (solid line) and in the strong-coupling
Fermi-surface-restircted
schemes (dashed line).  The sharp feaure in the
Fermi-surface-restricted calculations is the van Hove singularity.
Note that the electron-spin fluctuation coupling constant,
and hence the $T_c$'s, are not necessarily physical (see text).
\label{fig:RULN}}

\figure{Critical temperature $T_c$ in K vs. the number of holes
per site $n$ in the
two-dimensional, tight-binding band structure of Eq. (\ref{eq:BS})
with $t_1$ = 250 meV
and with $t =$ 0.45 to model the YBCO family.
The pairing is mediated by spin-fluctuations as described in
Ref. \cite{MBP}.
$T_c$ is computed both in the strong-coupling
exact (solid line) and in the strong-coupling
Fermi-surface-restircted
schemes (dashed line).  The sharp feaure in the
Fermi-surface-restricted calculations is the van Hove singularity.
\label{fig:MMP}}


\begin{references}
\bibitem{SF1}The RULN model is implemented as in Ref.
\cite{RULN} except that the electron-spin fluctuation coupling constant
$g^2$ = 1.6 eV$^2$ in the present calculations.  In the notation of
Ref. \cite{RULN}, $g^2 = (\pi / C N(0)) \,\bar{I}^2$, where
$C$ = 0.19 eV$^{-1}$ and $N(0)$ is the single spin density of
states at the Fermi surface.
\bibitem{MMPnote}The MMP model is implemented as in Ref.
\cite{Millis} except that the spectral function parameters have
been made temperature-dependent in accordance with the plots of the
$\rm YBa_2Cu_3O_7$ parameters displayed in Ref. \cite{MPT}.
The temperature dependences we use are
$(\xi / a)^2$ = 1111.2 K / (T + 120 K),
$\omega_{SF}$ = 0.380 T + 58.4 K, and
$\chi_Q$ = 42.6 eV$^{-1}$ - (0.0847 eV$^{-1}$K$^{-1}$) T,
where T is the temperature in K.  We also set the
electron-spin fluctuation coupling constant $g^2$ = 2.0 eV$^2$.
\bibitem{saturation}These large energy scales away from the
$(\pi,pi)$ points could be related to the fact that our calculated
critical temperatures saturate as function of coupling constant
at a larger value in the MMP model than in the RULN model.
\bibitem{Ein}For these calculations,
the electron-phonon coupling constant $g^2$
is chosen to be 0.01 eV$^2$ for the strong-coupling calculations and
0.005 eV$^2$ for the weak-coupling calculation.
Additionally, the Matsubara sum is cut-off at 500 meV, although
the results are insensitive to the precise value of this cut-off.
\bibitem{SF2}The exact calculations are done on a 32 x 32
lattice with the Matsubara sum truncated after the lowest 254
Matsubara frequencies;
the Fermi-surface-restricted $T_c$'s are computed with a
Matsubara cut-off of 250 meV.  The results presented are
insensitive to the choice of these parameters.
\bibitem{MMP1}The exact calculations are done on a 32 x 32
lattice with the Matsubara sum truncated after the lowest 512
Matsubara frequencies;
the Fermi-surface-restricted $T_c$'s are computed with a
Matsubara cut-off of 1000 meV.  The results presented are
insensitive to the choice of these parameters.
\bibitem{incom}By changing $\bf Q$ in this model from $(\pi,\pi)$
to the incommensurate value $(\pi,0.76\pi)$,
we have found a suppression of the d-wave $T_c$ as was
suggested in A. V. Chubukov, Phys. Rev. B {\bf 46}, 5588 (1992),
although the effect is only moderate (30 \%
for the one example we considered).
\end{references}
\end{document}